\documentclass[11pt,centertags,reqno]{amsart}

\usepackage[foot]{amsaddr}
\usepackage{latexsym}
\usepackage[english]{babel}
\usepackage[T1]{fontenc}
\usepackage[utf8]{inputenc}
\usepackage{natbib}

\usepackage{amssymb}
\usepackage{fancyhdr}
\usepackage{url}
\usepackage{hyperref}
\usepackage{verbatim}
\usepackage{leftidx}
\usepackage{color,graphicx}

\usepackage{mathrsfs, mathtools}
\usepackage{stmaryrd}

\usepackage{marvosym}
\mathtoolsset{showonlyrefs}

\usepackage{soul}
\usepackage{bm}

\usepackage{enumerate} 
\usepackage{enumitem} 

\numberwithin{equation}{section} \makeatletter
\renewcommand{\subsection}{\@startsection
{subsection}{2}{0mm}{\baselineskip}{-0.25cm}
{\normalfont\normalsize\bf}} \makeatother

\newtheorem{theorem}{Theorem}[section]

\theoremstyle{remark}
\newtheorem{remark}[theorem]{Remark}

\def \F {\mathcal F}

\def \R {\mathbb R}

\newcommand{\Rep}{\mathcal{R}}

\usepackage{orcidlink}

\begin{document}
	
\author[K.~Colaneri]{Katia Colaneri\orcidlink{0000-0003-3933-3788}}\address{Katia Colaneri, Department of Economics and Finance, University of Rome Tor Vergata, Via Columbia 2, 00133 Rome, Italy. \textrm{Corresponding author}.}\email{katia.colaneri@uniroma2.it}

\author[C.~Damian]{Camilla Damian\orcidlink{0000-0003-3687-3199}}\address{Camilla Damian, Institute of Statistics and Mathematical Methods in Economics, TU Wien, Wiedner Hauptstr. 7, 1040 Vienna, Austria}\email{camilla.damian@tuwien.ac.at}

\author[R.~Frey]{R\"{u}diger Frey\orcidlink{0000-0002-8402-4653}}\address{R\"{u}diger Frey, Institute for Statistics and Mathematics, Vienna University of Economics and Business, Welthandelpatz 1, 1020 Vienna, Austria}\email{rfrey@wu.ac.at}


\title[]{A filtering approach for statistical inference in a stochastic SIR model with an application to Covid-19 data}

	\maketitle

\begin{abstract}
In this paper, we consider a discrete-time stochastic SIR model, where the transmission rate and the true number of infectious individuals are random and unobservable. An advantage of this model is that it permits us to account for random fluctuations in infectiousness and for non-detected infections. However, a difficulty arises because statistical inference has to be done in a partial information setting.
We adopt a nested particle filtering approach to estimate the reproduction rate and the model parameters. As a case study, we apply our methodology  to Austrian  Covid-19 infection data. 
Moreover, we discuss forecasts and model tests.
\end{abstract}


\date{\today}

\section{Introduction}

Infectious diseases have key characteristics that simple epidemiological models may not be able to capture. To begin with, the infection or transmission rate may depend on several random factors, such as changes in the infectiousness of the virus, environmental conditions and seasonality, and eventually  policy measures. This suggests that the rate should be modelled as a stochastic process. Moreover, there is randomness in the transmission process of a disease, in particular if the population size is small.  Finally, many infections are not detected since medical tests might be necessary to confirm a diagnosis.  This implies that an analyst is not able to fully observe the components of the epidemiological system;  in particular, the transmission rate and the true number of infections are latent. 

To account for these characteristics, we develop a stochastic and partially observable SIR model with a random transmission rate. We first frame our model within the setting of a hidden Markov model (HMM). This consists of two components: a latent Markov process  (the so-called state process), and an observable process (the so-called measurement or observation process) which is affected by the state, see \cite{bib:elliott} and \cite{bib:churchill} for details. In our specific setup, the state process consists of the compartments of the SIR model and the of transmission rate. The observation process, on the other hand, is given by the number of newly confirmed cases. 
In a second step, we extend our framework to include quarantine measures. This is done by immediately removing the newly confirmed cases from the pool of infectious people. 
A stochastic epidemiological model with similar features (but without quarantine) has been considered in \citet{bib:stocks-et-al-20} for the spread of Rotavirus infections in Germany.

Statistical inference for this type of models is challenging because of noisy observations, as well as randomness and non-linearity in the transmission process.
To address this issue, \cite{bib:stocks-et-al-20} use a frequentist estimation methodology and compute  the MLE for the model parameters via  the \emph{iterated filtering} approach   of \cite{bib:ioanides-et-al-16}.  In this paper, instead, we analyse the problem using a Bayesian approach. Our objective is to approximate the posterior distribution of the state variables and of the model parameters. For this purpose, we rely on sophisticated techniques from stochastic filtering and adapt the nested particle filter  of \cite{bib:crisan-miguez-18} to our setup. The ensuing algorithm is recursive and can be implemented efficiently. Simulation experiments demonstrate the good performance of this methodology.



Our approach has many advantages. First, because of its Bayesian nature, it is easily adapted if additional sources of information -- for example, in our context, sewage data -- become available. In a similar spirit, expert opinions on the state of the epidemiological system are readily incorporated into prior distributions.  
Second, our methodology is ideally suited for forecasting  of  future infection numbers. This is an important application of epidemiological models, as these forecasts are used to gauge potential stress for the health system and to inform  decisions on containment measures. In our context, forecasts are based on the \emph{predictive posterior distribution}, so that  parameter uncertainty is taken into account naturally, making the procedure robust. 
Last, our approach makes it possible to conduct formal goodness-of-fit tests for the predictive distribution.  

As a case study, we apply the version of the model with quarantine to Covid-19 infection data from Austria. A comparison with the official estimates of the Austrian health agency AGES  shows that our approach produces qualitatively similar results for the {\em effective reproduction number}. 
We also discuss an application of our goodness-of-fit tests  in the context of Austrian Covid-19 data, which  provides further support for our methodology. Summarizing, our results show that  statistical inference for an elaborated epidemiological model with partial information is indeed feasible, if one uses recently-developed tools from stochastic filtering.

We continue with a discussion of the relevant literature. To the best of our knowledge, there are  only  few contributions that treat statistical inference for epidemiological models as an estimation problem under partial information.
\cite{bib:hasan-et-al-nature} use the extended Kalman filter (EKF) in an SIR model with additional Gaussian noise to analyse  Scandinavian Covid-19 data. However, their model dynamics are somewhat implausible; in particular, the effective reproduction number is modelled as a random walk with Gaussian increments and can therefore become negative. Moreover, the SIR model is a non-linear system and the EKF linearizes these dynamics in a somewhat ad-hoc fashion, so that optimality and stability of the estimates cannot be guaranteed (see e.g. \cite{bib:budhiraja-chen-lee-07}). However, the EKF is comparatively simple and it  often works well when  non-linearities are small.

Besides \cite{bib:stocks-et-al-20}, a few other interesting papers are \cite{bib:sun-et-al} and \cite{corbella2022inferring}. 
\cite{bib:sun-et-al} study parameter estimation for three partially observed dynamical models for the evolution of biological, ecological or environmental processes; namely,  a deterministic model, a Markov-chain-based model and a model described by stochastic differential equations. They address the inference problem using approximate Bayesian computing.   
In \cite{corbella2022inferring}, an epidemiological system is described via a state-space model, where randomness is used to address relatively rare severe events. The authors use multiple dependent datasets and develop algorithms for parameter inference based on a pseudo-marginal approach. 

The rest of the paper is organized as follows. In Section~\ref{sec:model} we introduce the model; in Section~\ref{sec:methodology} we describe the nested particle filter; in Section~\ref{sec:simulation} we present results for simulated data; Section~\ref{sec:empirical} is concerned with applications to Austrian infection data; finally in Section~\ref{sec:predictions} we discuss forecasts and model tests.
Appendices \ref{app:particle-filter}, \ref{app:simulation} and \ref{app:data_parameters} collect, respectively, some comments on nested particle filter, inputs for the simulation study and inputs for the real data analysis.

\section{Model specification}\label{sec:model}

In the sequel, we introduce a stochastic version of the standard SIR model, where both the actual number of infected people and the infection rate are random and unobservable. Since we will test our methodology on Covid-19 data, our model specifications account for some special features of this context, such as testing and quarantine regulations. We describe the model in two steps. In the first step (Section \ref{sec:step1}), we consider a simpler case in which quarantine is not taken into account, and develop a discrete-time HMM for the epidemiological system. In the second step, we include quarantine. This generates an additional dependence channel between the state variable and the observations that we explain better in Section \ref{sec:step2}. 

\subsubsection*{Notation.} We begin by introducing key variables of the model. First, since infection numbers are usually reported on a daily or a weekly basis, we work in  a discrete-time setting with time points $(0=t_0,t_1, \dots, t_n, \dots)$ (in the data analysis, we assume that $t_n - t_{n-1}$ is one day).
We consider a population with $N$ individuals and we assume, for simplicity, that the population size stays constant over time.\footnote{This assumption approximates the case where the observation period is short and the number of deaths due to infections is small compared to the population size.}  We then let
\begin{itemize}
\item $S_n$ be the number of susceptible individuals at time $t_n$;
\item $I_n$ be the number of infectious persons at time $t_n$ who can generate new infections in the period  $[t_n, t_{n+1})$;
\item $I_n^+$ be the number of individuals who get infected in  $[t_{n}, t_{n+1})$;
\item $P_n$  be the number of newly \emph{reported} infections (such as positive tests) in the interval $[t_{n}, t_{n+1})$, where we assume that testing starts at $t_1$;
\item $I_n^-$  be the number of individuals  who were infectious at $t_{n}$ but are removed from $I_n$ over $[t_n, t_{n+1})$, for instance since they recovered or, for certain diseases, are in quarantine;
\item $R_n$  be the number of so-called \emph{removed} individuals; that is, people who are either immune or in quarantine at time $t_n$;
\item$\Psi_n$ be the logarithmic transmission rate; roughly speaking, $\beta_n:=\exp(\Psi_n)$ is the expected number of people that are infected by a single infectious person in the period $[t_n, t_{n+1})$. 
\end{itemize}
A process is indicated by capital letters without time index, e.g. $S$  denotes the discrete-time process $(S_n)_{n = 0, 1, \dots}$. 
We will also use the notation $P_{1:n}$ to indicate the the history of the process up to $n$, that is the sequence of random variables $P_1, \dots, P_n$.  
Finally, we adopt the usual convention that upper case letters are random variables and lower case variables are data points or samples.

\subsection{Step 1: an HMM for epidemics without quarantine}\label{sec:step1}
In what follows, we deal with a set of variables, called the \emph{stock} variables, that represent the number of people in each compartment at time $t_n$, given by $(S_n, I_n, R_n)$. These, together with the logarithmic infection rate $\Psi_n$, form the unobservable state. There is also a set of so-called \emph{flow} variables, given by $(I^+_n, I^-_n, P_n)$, that are used to represent the changes in the stock variables from $t_{n}$ to $t_{n+1}$. Among flow variables,  $I^+_n$ and $I^-_n$ are latent, whereas $P_n$ provides the observation.  

\subsubsection{The state variables}
In this section we describe the evolution of the state of system, namely the dynamics of processes $(S, I, R)$ and $\Psi$ .  Throughout, we fix a distribution for $I_0$,  $R_0$ and $\Psi_0$.
We begin with the logarithmic transmission rate. We assume that $\Psi$ follows  a first-order autoregressive process with the dynamics
\begin{equation}\label{eq:dynamics-Psi}
\Psi_{n} =  \Psi_{n-1} + \kappa (\mu -\Psi_{n-1})  + \sigma Z_{n-1}\,, \quad n = 1,2,\dots
\end{equation}
for a sequence of independent standard normal random variables $\{Z_n\}_{n=0,1, \dots}$ and parameters $\kappa, \sigma >0$ and $\mu \in \R$. 


Next, we introduce the dynamics of $S$, $I$ and $R$. By definition, the number of susceptible people satisfies $$S_n = N - I_n - R_n, \quad n =0,1,\dots,$$so that $S_n$ can be identified from  $I_n$ and $R_n$ (for this reason, we can omit $S_n$ in the set of state variables).  The process $I$ evolves according to 
$$I_n=I_{n-1}+I^+_{n-1}-I^-_{n-1}, \quad n=1,2,\dots \ .$$
The new infections $I_{n}^+$ are modelled as a Poisson random variable,
\begin{equation}\label{eq:def-beta}
I_{n}^+ \sim \mbox{Poisson} (\lambda_{n}) \; \text{ with } \lambda_{n} = \beta_{n} I_{n} \frac{S_{n}}{N}, \quad n=0,1,\dots,
\end{equation}
where $\beta_{n}=\exp(\Psi_{n})$. The model states that the expected number of new infections is proportional to the number of infectious people and to the fraction of susceptible individuals in the whole population.  The proportionality factor $\beta_{n}$ gives the average number of people that are infected in $[t_{n}, t_{n+1})$ by one infectious person in a population where everyone is susceptible, hence it is named \emph{infection} or \emph{transmission} rate. 

\begin{remark}
We now briefly comment on our assumptions on the transmission model.
\begin{itemize}
\item[(i)]It is natural to assume that, for small $t_{n+1}-t_{n}$, the quantity  $\beta_{n} \frac{I_{n}}{N}$ is small. In that case, we can interpret it as the probability that a susceptible person at time $t_{n}$ gets infected over the interval $[t_{n}, t_{n+1})$. If, moreover, infection events are assumed to be independent across susceptible individuals, and the susceptible population $S_{n}$ is large, then we can use the Poisson approximation to justify the model \eqref{eq:def-beta}.
\item[(ii)]To motivate the assumption for the infection rate dynamics \eqref{eq:dynamics-Psi}, we observe that this process is stationary and mean-reverting around the value $\mu$, which is a common behaviour for both  endemic and pandemic diseases. In fact, the infection rate of an endemic disease is stationary by definition, whereas, for pandemics, stationarity is often enforced by policy measures. For instance, in the pandemic phase of Covid-19, many European governments tightened containment rules in periods of high infection numbers (corresponding to high values of the reproduction index $\Rep_n$, see below) and loosened measures after infection numbers had fallen to more sustainable levels.
\end{itemize}
\end{remark}

Infectious people who are not detected move to the removed state upon recovery from the infection. Thus, at the end of the time interval $[t_{n}, t_{n+1})$, the number of infectious people is reduced by $$ I_{n}^- = \gamma I_{n}, \quad n=0,1,2,\dots, $$where $\gamma > 0$ is the inverse of the average time a non-detected individual stays infectious. 
Finally, we assume that 
$$R_{n}=R_{n-1}+I_{n-1}^- -\delta R_{n-1}, \quad n =1,2,\dots,$$
where the parameter $\delta>0$ is such that $1/\delta$ is the average time an infected person enjoys immunity. In other words, a removed person looses immunity and becomes susceptible again at rate $\delta$.  For instance,  $\delta \sim \frac{1}{200}$ says that people who recovered from the virus on average do not get infected again for about 200 time units (days in our case). 
Summarizing, the dynamics of the system are as follows. For $n =1, 2, \dots$, 
\begin{equation}\label{eq:dynamics-1}
\left \{ \begin{split}
\Psi_n&=\Psi_{n-1}+ \kappa (\mu -\Psi_{n-1})  + \sigma Z_{n-1}\\
I_n&=I_{n-1}+I_{n-1}^+- I^-_{n-1}\\
R_{n}&=R_{n-1}+ I_{n-1}^- -\delta R_{n-1}\,,
\end{split}\right.
\end{equation}
where $I_{n}^+\sim \mbox{Poisson}\big(\beta_{n} \frac{I_{n}}{N} S_{n}\big)$ and $I_n^-=\gamma I_{n}$ and $Z_{n}\sim N(0,1)$. 
It is clear from \eqref{eq:dynamics-1} and \eqref{eq:dynamics-Psi} that the distribution of the triple $(I_n,R_n,\Psi_n)$ can be described in terms of a transition kernel that depends only on  $(I_{n-1},R_{n-1},\Psi_{n-1})$, therefore forming a discrete-time  Markov chain. 

\subsubsection{The observations}
The true number of infectious people is unknown, as  this quantity includes asymptomatic infections or infected individuals who have not (yet) taken  a test. Since infections are random and unobservable, we cannot observe the infection rate $\beta_n$ (or equivalently its logarithm,  $\Psi_n$).  At any time $n=1,2,\dots$,  the available information is thus provided by the number of newly reported cases $P_n$, whereas all other variables that are used to identify the state are latent. 
To describe the dynamics of new cases, we assume that an infectious person at time $t_{n}$ is detected in the interval $[t_{n}, t_{n+1})$ with probability $q\in [0,1]$. The parameter $q$ accounts for the availability and reliability of tests and/or for the intensity of public screening programs.
We assume that testing occurs independently across infected people and starts at time $t_1$. In that case, the conditional distribution of positive tests at time $t_n$, given the number $I_{n}$ of infectious people at time $t_{n}$, is binomial with parameters $I_{n}$ and $q$. That is to say, 
$$P_{n}  \sim \mbox{Binomial}\left(\lfloor I_{n}\rfloor, q\right),\quad n =1,2,\dots.$$
where $\lfloor \cdot \rfloor$ denotes the floor function. Formally, at any time $t_n$, the available information can be described by the history $P_{1:n}$. 

Note that other data sources, such as results from sewage screening or sentinel systems, are easily integrated into our approach, provided that the conditional distribution of these variables given $I_n$ is known (however, this is left for future research).

\subsection{Step 2: an extension with quarantine}\label{sec:step2}
In this section, we consider an extension of the model in Step 1, where quarantine is introduced. This, in turn, implies a modification in the dynamics of removed individuals; that is, we assume that a person who tests positively is immediately removed from the pool of infectious people to reflect quarantine measures or self-isolation precautions. In addition, infectious people who are not detected move to the removed state upon recovery from the infection, as before.
Thus, for any $n=0,1,2,\dots$, the number of infectious people is reduced by $$I_n^- = P_n + \gamma I_{n},$$
where we set $P_0=0$ as we assumed that testing starts at time $t_1$.\footnote{Note that we use the same notation $I_n^-$ to indicate the number individuals who are removed from the pool of infectious with and without quarantine, but their dynamics are different.} In summary, the dynamics of the system are as follows. 
For $n = 1,2,\dots$,  
\begin{equation}\label{eq:dynamics_2}
\left \{ \begin{split}
\Psi_n&=\Psi_{n-1}+ \kappa (\mu -\Psi_{n-1})  + \sigma Z_{n-1}\\
I_n&=I_{n-1}+I_{n-1}^+- I^-_{n-1}\\
R_{n}&=R_{n-1}+ I_{n-1}^- -\delta R_{n-1}\,,
\end{split}\right.
\end{equation}
where $I_{n}^+\sim \mbox{Poisson}\big(\beta_{n} \frac{I_{n}}{N} S_{n}\big)$,  $I_n^-=P_n+\gamma I_{n},$ $P_{n}  \sim \mbox{Binomial}\left(\lfloor I_{n}\rfloor, q\right)$ and $Z_n \sim N(0,1)$. As before, the observations are given by the history of confirmed cases,  $P_{1:n}$ for every $n=1,2, \dots$ .

In the following statistical analysis we use the property that the triple $(I, R, \Psi)$ is conditionally Markovian, given the observations, which is due to the fact that $(I,R,\Psi,P)$ is Markovian. Moreover, note that the state variables in the model with quarantine cannot be described independently of the observations, which directly appear in their dynamics. Therefore, this model no longer falls into the class of standard HMMs.

\subsection*{Reproduction rate.} The effective reproduction rate $\Rep_n$ is an index that measures  the number of individuals who are infected, on average, by a specific infectious person, given the state of the pandemic system at time $t_n$. 
To  identify $\Rep_n$ in the model with quarantine,\footnote{For the version of the model without quarantine it is known that $\mathcal{R}_n\approx \frac{\beta_n}{\gamma}\frac{S_n}{N}$.} note first that an infected individual transmits the disease, on average, to $\beta_n S_n/N$ people per day. Moreover,  the time that elapses before an infected person transits to the compartment of removed individuals, $R$, is the minimum of $ \tau^{\text{rec}}$ (the time up to recovery) and $ \tau^{\text{quar}}$ (the time until the  person tests positively and is put into quarantine). Under our model dynamics, $\tau^{\text{rec}}$ and $\tau^{\text{quar}}$ have a geometric distribution with parameter $\gamma$  and $q$, respectively. Moreover they are independent, so that  $\min\{ \tau^{\text{rec}}, \tau^{\text{quar}}  \}$ follows  a geometric distribution with parameter $\gamma+ q - \gamma q \approx \gamma + q$.  Hence, the expected time up to removal satisfies
$ \mathbb{E}(\min\{ \tau^{\text{rec}}, \tau^{\text{quar}}  \}) \approx (\gamma + q)^{-1} \, , $
and
\begin{equation}\label{eq:Rep-n}
\Rep_n \approx \frac{\beta_n}{\gamma + q} \frac{S_n}{N} \, .
\end{equation}

\section{Statistical Methodology}\label{sec:methodology}
When discussing statistical inference for the model described in Section~\ref{sec:model}, we can distinguish two problems: the \textit{filtering} problem and the \textit{parameter estimation} problem. The filtering problem is concerned with inferring the state process -- in our case, the triple $(I, R, \Psi)$ -- conditional on the available observations and on a given parameter vector $\bm{\Theta}$. The parameter estimation problem, on the other hand, is concerned with the case in which parameters values are also unknown. A straightforward way to estimate state and parameters jointly within a bootstrap filter would be to augment the state space so that it includes a parameter vector as a constant-in-time variable; however, this implies that the parameter space is explored only in the initialization step of the algorithm, making it destined to degenerate, see e.g. \cite{bib:kantas}. It is thus important to use a methodology that periodically reintroduces diversity in the parameter space.

Having this in mind, in this paper we chose to adapt the nested particle filtering (NPF) algorithm of \citet{bib:crisan-miguez-18} to track the posterior distribution of the (static) unknown parameters in our model, as well as the joint posterior distribution of parameter and state variables, in a recursive fashion.  More specifically, the NPF algorithm consists of two nested layers of particle filters: an ``outer'' filter, which approximates the posterior of $\bm{\Theta}$ given the observations, and a set of ``inner'' filters, each corresponding to a sample generated in the outer layer and yielding an approximation of the posterior measure of the state conditional on both the observations and the given sample of $\bm{\Theta}$. In this section, we briefly describe this methodology within our framework, starting with the standard bootstrap filter (conditional on a given parameter sample) for the state which are batched to build the inner layer of the NPF algorithm.

\subsection{Inner (State) Filter}\label{subsec:inner}
The filtering problem is concerned with inferring the state process -- in our case, the triple $(I, R, \Psi)$ -- conditional on the available observations and on a given parameter vector $\bm{\Theta}$. Assuming that both the transition density of the state process and the conditional density, at each discrete-time instance, of observation given signal and parameters are known, the goal reduces to tracking the posterior probability distribution of the state and it can be accomplished by standard particle (bootstrap) filtering (see, e.g. \cite{bib:gordon}). In what follows, we provide a schematic representation of the bootstrap algorithm in our context.
\begin{enumerate}\setcounter{enumi}{0}
\item \textit{Initialization ($n = 0$)}: draw i.i.d. state particles $\left(i_0^{(m)}, r_0^{(m)}, \psi_0^{(m)}\right)$, $m = 1, \dots, M$ from the prior distribution.
\item \textit{Recursive step (from $n - 1$ to $n$)} 
Given a total of $M$ state particles (Monte Carlo samples) available at time $n - 1$ and the new observation $p_n$, at time $n$:
\begin{itemize}
\item[(a)] \textit{Propagate}: propagate $\left(i_{n - 1}^{(m)}, r_{n - 1}^{(m)}, \psi_{n - 1}^{(m)}\right)$ to $\left(\bar{i}_n^{(m)},  \bar{r}_n^{(m)}, \bar{\psi}_n^{(m)}\right)$, $m = 1, \dots, M$, according to Equation \ref{eq:dynamics-1} (for the no-quarantine model) or Equation~\eqref{eq:dynamics_2} (for the quarantine model).
\item[(b)] \textit{Compute normalized weights}: compute particle weights proportional to the binomial likelihood; i.e. 
we define $a^{(m)}_n = \binom{\bar{i}_n^{(m)}}{p_n} q^{p_n} (1 - q)^{\bar{i}_n^{(m)} - p_n}$, and then the normalized weights are given by  $\bar{w}^{(m)}_n = \frac{a^{(m)}_n}{\sum_{m = 1}^M a^{(m)}_n}$.
\item[(c)] \textit{Resample}: for $m = 1, \dots, M$, let $\left(i_n^{(m)}, r_n^{(m)}, \psi_n^{(m)}\right) = \left(\bar{i}_n^{(j)},  \bar{r}_n^{(j)}, \bar{\psi}_n^{(j)}\right)$ with probability $\bar{w}_n^{(j)}$, $j \in \{1, \dots, M\}$.
\end{itemize}
\end{enumerate}

\subsection{Parameter Estimation via Nested Particle Filtering}  
When it comes to parameter estimation in our setup, we can distinguish two sets of parameters influencing the epidemiological system. On the one hand, we have the triple $\bm{\theta} = (\kappa, \sigma,\mu)^\top$, which governs the dynamics of the logarithmic infection rate. On the other hand, we have the rates $q$, $\gamma$ and $\delta$. It is worth noticing that our observation process consists only of the number of reported cases and that, due to such limitations in the nature and in the length of the observation time series, we will not be able to estimate all model parameters with reasonable accuracy and we will thus resort to fixing the rates exogenously.\footnote{A formal analysis of this issue is given in \cite{bib:stocks-supplementary}. They show that, for a deterministic SIR model in its endemic (stationary) state, only the infection rate $\beta$ can be estimated from infection data, while the other SIR parameters have to be estimated from other data sources.} That is, we assume that the parameters $q$, $\gamma$ and $\delta$ are derived from other data (for instance, different medical data, sewage data, as well as the results of other statistical studies) and hence represent a fixed input of our model. As an example of a study reporting calibrated detection rates for Austria specifically, we refer the reader to \cite{rippinger2021evaluation}.

\subsubsection*{Nested Particle Filtering Algorithm} Next we describe the NPF of  \citet{bib:crisan-miguez-18}, which is used for estimating the posterior distribution of the state variables and of the unknown parameters. 
To simplify notation, in the following schematic description of the algorithm we denote by  $X_n = (I_n, R_n, \Psi_n)^\top$, $n = 0, 1, \dots$, the state variables.

\begin{enumerate}
\item  {\em Initialisation.}
\begin{itemize}
\item Draw $K$ independent and identically distributed samples of the parameter vector, $\theta_0^{(k)}$, $k=1, \dots, K$, from a prior distribution.
\item For each parameter configuration $\theta_0^{(k)}$, $k=1, \dots, K$, draw $M$ independent and identically distributed samples of the state variables, $x_0^{(k, m)}$ , $m = 1, \dots, M$, from a prior distribution, so that the total number of particles in the signal space is $K \cdot M$.
\end{itemize}
\item  {\em Recursive step.} Let, for $n \ge 1$,  $(\theta_{n-1}^{(k)}, x^{(k, m)}_{n-1})$ be the set of available samples at time $n-1$:
\begin{enumerate}
\item {\em Jittering and state propagation.} For each $k = 1, \dots, K$:
 \begin{enumerate}
 \item Draw parameter particles $\bar \theta^{(k)}_n=(\bar \kappa^{(k)}_n, \bar \sigma^{(k)}_n, \bar \mu^{(k)}_n)^\top$ by perturbing the available sample using a jittering kernel. We jitter each parameter using a Gaussian distribution truncated on its support with means  $\kappa^{(k)}_{n-1}$, $\sigma^{(k)}_{n-1}$ and $\mu^{(k)}_{n-1}$, respectively, and variances $\epsilon_{\kappa}$, $\epsilon_\sigma$, and $\epsilon_\mu$, respectively.
\item Conditional on each parameter vector $\bar \theta^{(k)}_n$, perform a inner bootstrap filtering step as described in Section~\ref{subsec:inner}. For each $m = 1, \dots, M$, propagate $x^{(k, m)}_{n-1}$ to $\bar{x}^{(k, m)}_{n}$ and compute the corresponding (binomial) likelihood $a^{(k, m)}_n$ and normalized likelihood weight $\bar w^{(k, m)}_n = \frac{a^{(k, m)}_n}{\sum_{m=1}^M a^{(k, m)}_n}$. For the given $\bar \theta^{(k)}_n$, update the corresponding state particles in a standard way (weigh and resample with replacement by means of the multinomial resampling algorithm) to obtain a new set in the signal space, $\tilde{x}^{(k,m)}_{n}$.
\end{enumerate}
\item {\em Resample parameter particles with replacement.} Compute the approximate the likelihood of each $\bar \theta^{(k)}_n$ and the likelihood weight as $w^{(k)}_n = \frac{\sum_{m = 1}^M a^{(k,m)}_n}{\sum_{k = 1}^K \sum_{m=1}^M a^{(k,m)}_n}$. For each $k=1, \dots, K$, set $(\theta^{(k)}_n, x^{(k, 1)}_n, \dots, x^{(k, M)}_n)=(\bar \theta^{(i)}_n, \tilde x^{(i, 1)}_n, \dots, \tilde x^{(i,M)}_n)$ with probability $w_n^{(i)}$, $i \in \{1, \dots, K\}$.
\item {\em Approximate the posterior measure.} Approximate the posterior distribution of the parameters by  $\mu_n\sim \frac{1}{K} \sum_{k=1}^K \delta_{\theta^{(k)}_n}$, where $\delta_{\theta}$ denotes the Dirac measure at point $\theta$.
\end{enumerate}
\end{enumerate}

It is important to note that the first task performed in a given iteration of the algorithm is the \textit{jittering} of existing parameter particles in order to restore diversity in the sample, as such diversity might have been greatly reduced due to a previously occurred resampling step.
For further computational details, the reader is referred to Appendix \ref{app:particle-filter}; moreover, details on inputs and prior choices for the simulation (resp. the data analysis) are given in Appendix~\ref{app:simulation} (resp. Appendix~\ref{app:data_parameters}).

\section{Simulation results}\label{sec:simulation}

\begin{figure}[h]
  \centering
  \includegraphics[scale = 0.38]{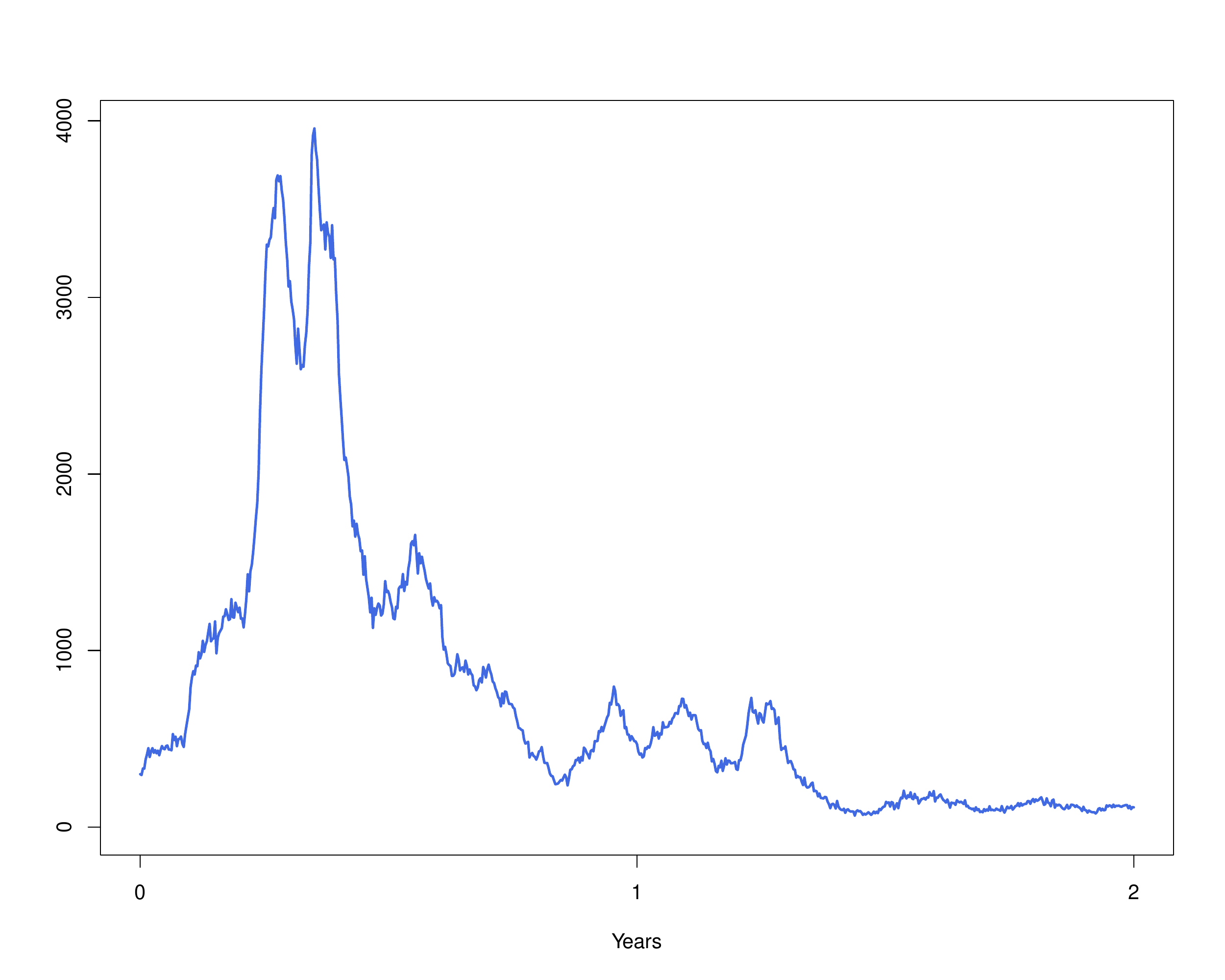}
  \caption{\footnotesize Simulated trajectory of the number of positive tests (observation process).}\label{fig:simulated_processes}
\end{figure}

The goal of this section is to test the previously-described nested particle filtering approach for the model with quaratine restrictions on simulated data that mimic typical features of Austrian Covid-19 data -- the {\em true} parameters used to generate state and observation time series are given in Appendix \ref{app:simulation}. We assumed that the probability of a positive test, the average duration of the illness and the average natural immunity period (i.e. $q, 1/\gamma, 1/\delta$) are given and we fixed them consistently with the values reported by e.g.~the Austrian Ministry of Health, see \cite{bib:richter-et-al-AGES}.
We ran the simulation for a period of two years ($731$ days), which is roughly consistent with the length of the real time series used for the application in Section~\ref{sec:empirical}. Figure~\ref{fig:simulated_processes}, depicts a simulated trajectory of the observation sequence $P_n$, $n = 1, \dots, 731$. This path shows qualitative properties that are similar to real Covid-19 infection data: for instance, our model naturally generates waves of infections. Note that, towards the end of the simulation period, the number of positive tests becomes low, which affects the accuracy of the estimation of the infection rate and the effective reproduction rate (see also Figure~\ref{fig:ERR} below).

We have used the nested particle filtering approach described in Section \ref{sec:methodology} to estimate the number of infections, the infection rate  $(\beta_n)_{n \geq 1}$, the reproduction rate $(\Rep_n)_{n \geq 1}$ and the model parameters. The results discussed below have been averaged over 50 independent rounds of the algorithm.
Figure~\ref{fig:ERR} displays the true (in black) and filtered (in magenta) trajectory of the effective reproduction rate $(\Rep_n)_{n \geq 1}$\footnote{We recommend the reader to use a color (or screen) version of this plot for a better understanding.}.  This plot suggests that the true trajectory exhibits higher variance than the filtered one: the filter generally captures the trend of the signal process quite well, but it is not able to detect small movements in a short time interval. Moreover, as mentioned previously, the amount of positive test is quite low towards the end of the simulation period and thus, as the observation process becomes less informative, the accuracy of the filter decreases.

\begin{figure}[h]
  \centering
  \includegraphics[scale = 0.42]{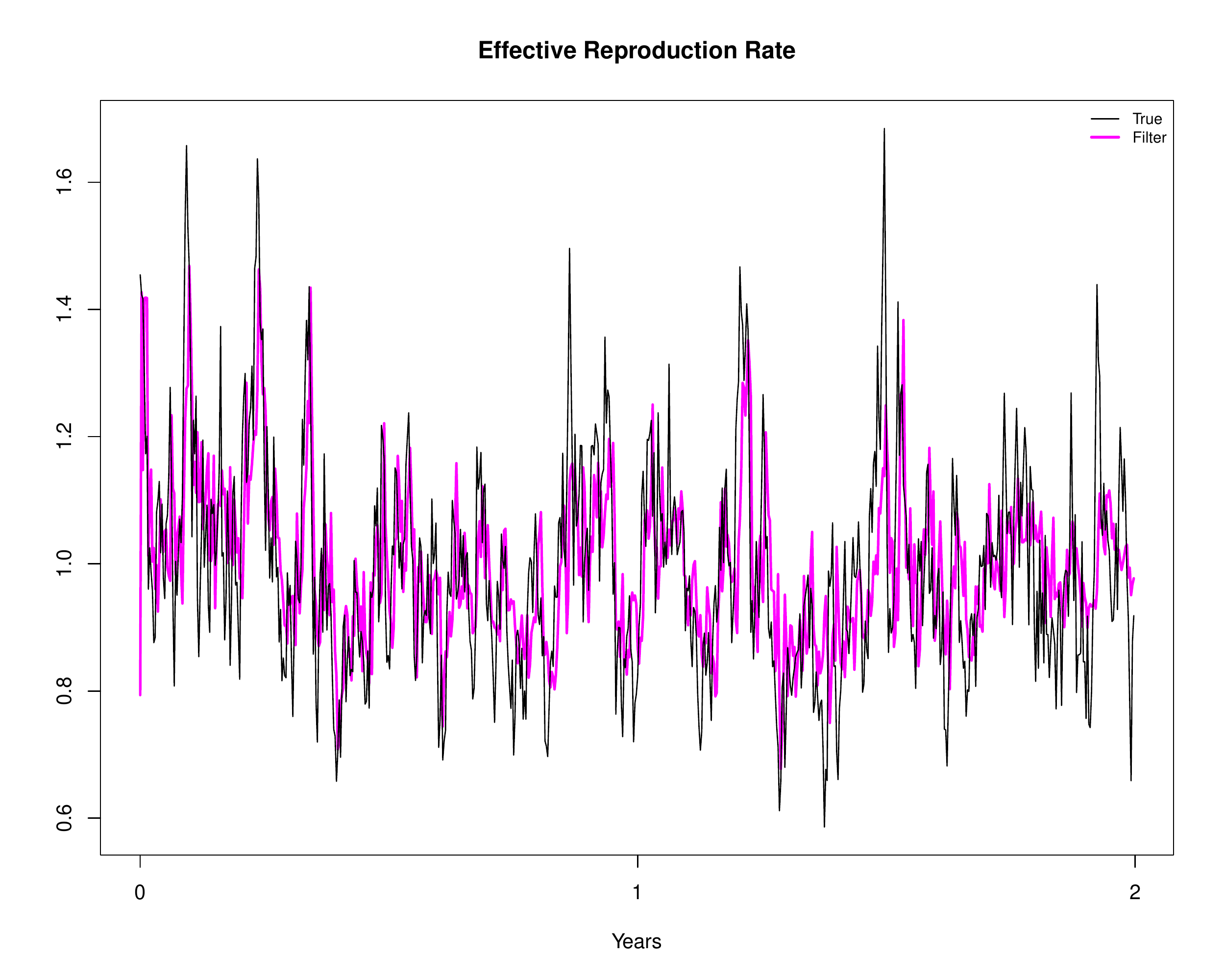}
  \caption{\footnotesize True (black) and filtered (magenta) trajectory of the effective reproduction rate.}\label{fig:ERR}
\end{figure}

\begin{figure}[h]
  \centering
  \includegraphics[scale= 0.42]{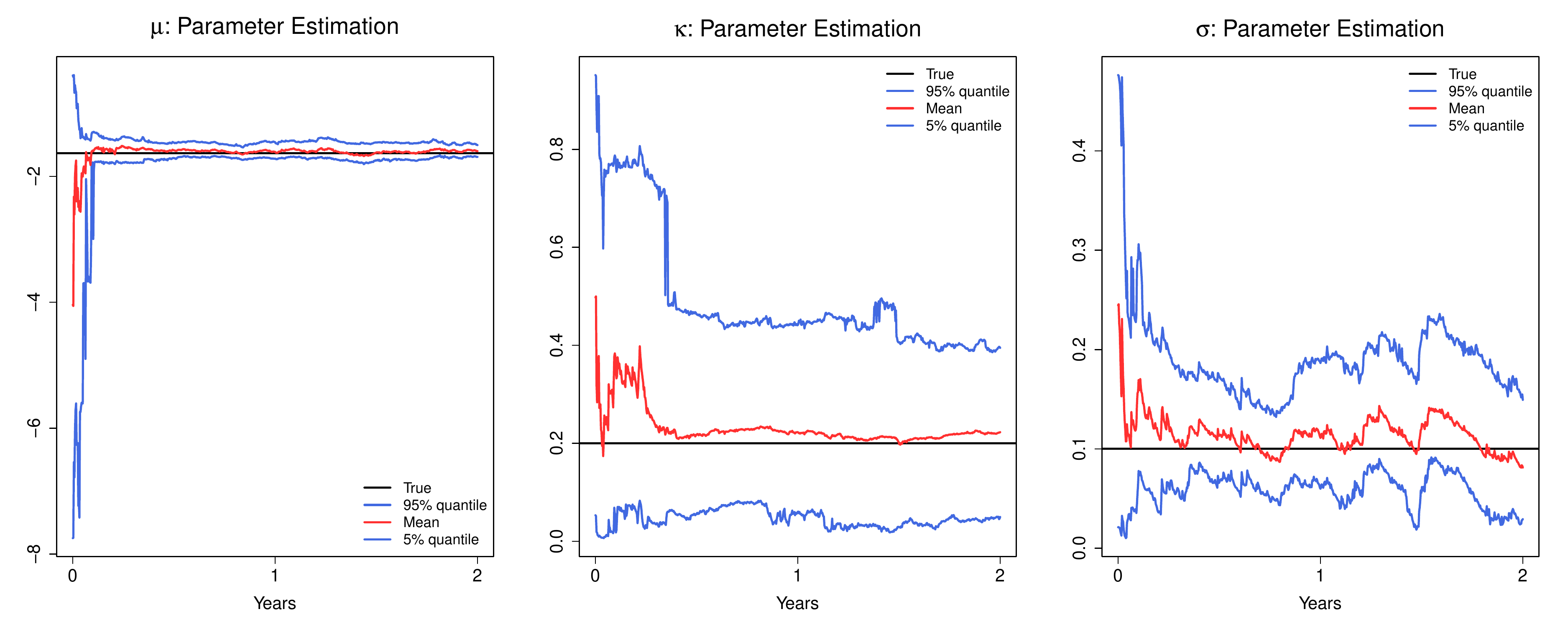}
\caption{\footnotesize Posterior estimates for $\mu$, $\kappa$, and $\sigma$. The black line corresponds to the true value of the parameter, the two blue lines to the 5\%- and 95\%-quantile of the posterior distribution (as obtained by the nested particle filter) and the red line to its mean.}\label{fig:kappa}
\end{figure}

Next, we discuss the estimation of the three unknown model parameters: $\kappa$, $\sigma$ and $\mu$ (see the dynamics of $\Psi_n = \log(\beta_n)$ given in equation~\eqref{eq:dynamics-Psi}). Their posterior distributions are obtained through the nested particle filtering algorithm and visualized in Figure~\ref{fig:kappa}. In each panel, the black line corresponds to the true value of the parameter, the two blue lines to the $5\%$- and $95\%$-quantile of the posterior distribution (as obtained by the nested particle filter) and the red line to its mean.

\begin{figure}[h]
  \centering
  \includegraphics[scale = 0.62]{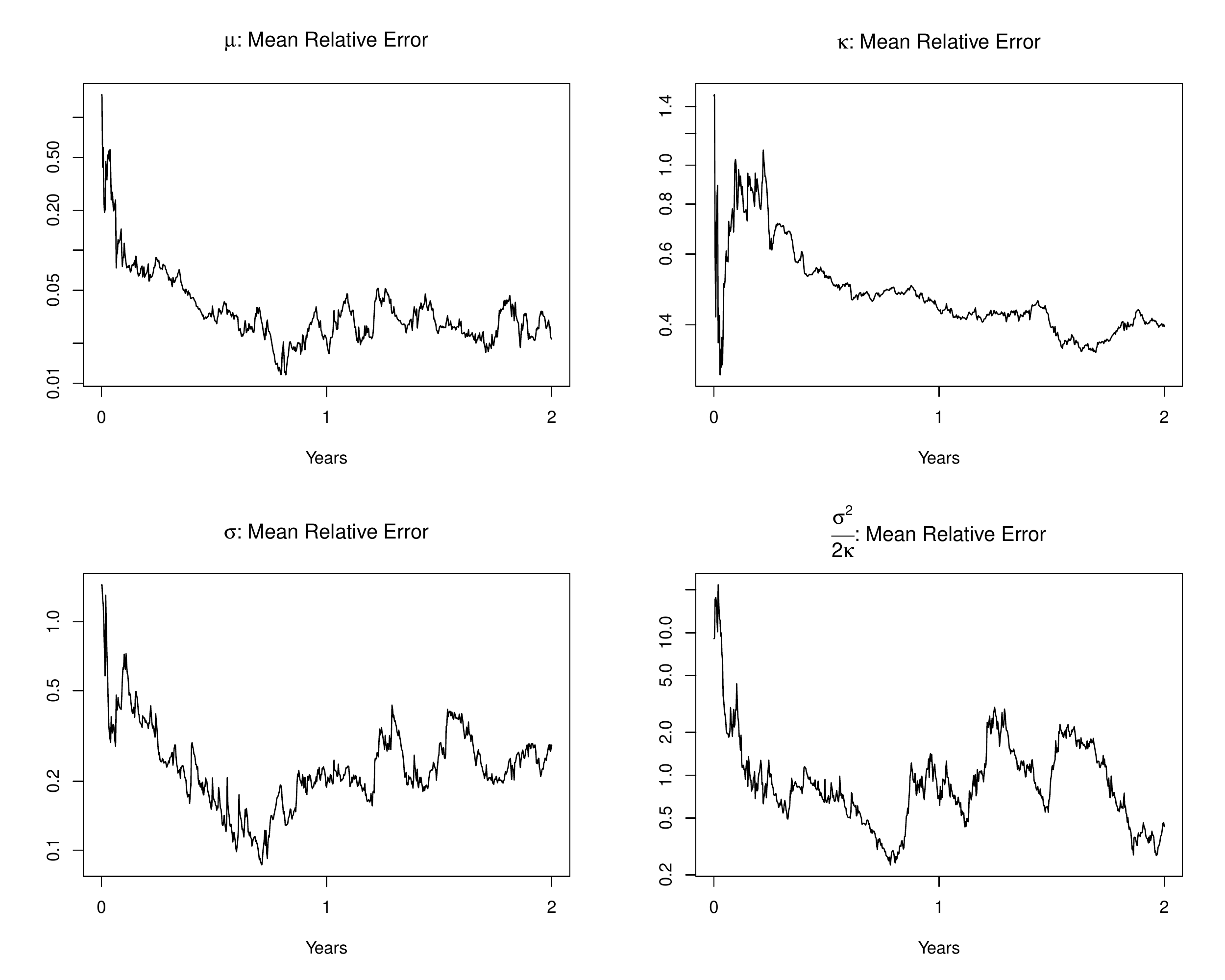}
 \caption{\footnotesize Mean relative errors for $\mu$ (top left panel), $\sigma^2/2\kappa$ (top right panel), $\kappa$ (bottom left panel) and $\sigma$ (bottom right panel), on a logarithmic scale.}
\label{fig:errors}
\end{figure}

We can observe how the posterior-mean estimate of $\mu$ quickly settles around the true value, while $\kappa$ and $\sigma$ seem to be more difficult to estimate.  We attribute this difficulty to a couple of reasons. First, we have relatively few observations, corresponding to two years\footnote{We decided to run our algorithm for two years only to be roughly consistent with the amount of data we used in our case study for the Austrian Covid-19 data.}. Second, we considered a quite small support for these parameters, within which it might be more challenging to further discriminate between plausible values in a short time span. Nevertheless, the algorithm is able to detect parameter magnitudes quite rapidly. Moreover, note that there might be a `compensatory' effect at play between posterior-mean estimates for $\sigma$ and $\kappa$, since the long-run variance of $\Psi_n$ is  $\sigma^2/2\kappa$, and their distict impact on state and observations might be decoupled by the algorithm only over a longer time horizon. The mean relative errors between true values and posterior-mean estimates, averaged over the 50 independent runs of the algorithm, are plotted in Figure \ref{fig:errors} and they appear consistent with our considerations: in particular, one can compare the behavior of parameter estimation errors for $\kappa$ and $\sigma$ with those for the ratio $\sigma^2/2\kappa$. Estimation errors for $\mu$ decrease quickly, as expected.

Finally, we carried out  robustness checks to ensure that small changes in the parameters $q$ and $\gamma$ do not affect our estimates too strongly, and we observed that the filtered effective reproduction rate corresponding to slightly-varied values of these parameters presents very similar qualitative and quantitative characteristics.

\section{Empirical results}\label{sec:empirical}

\begin{figure}
  \centering
  \includegraphics[scale = 0.42]{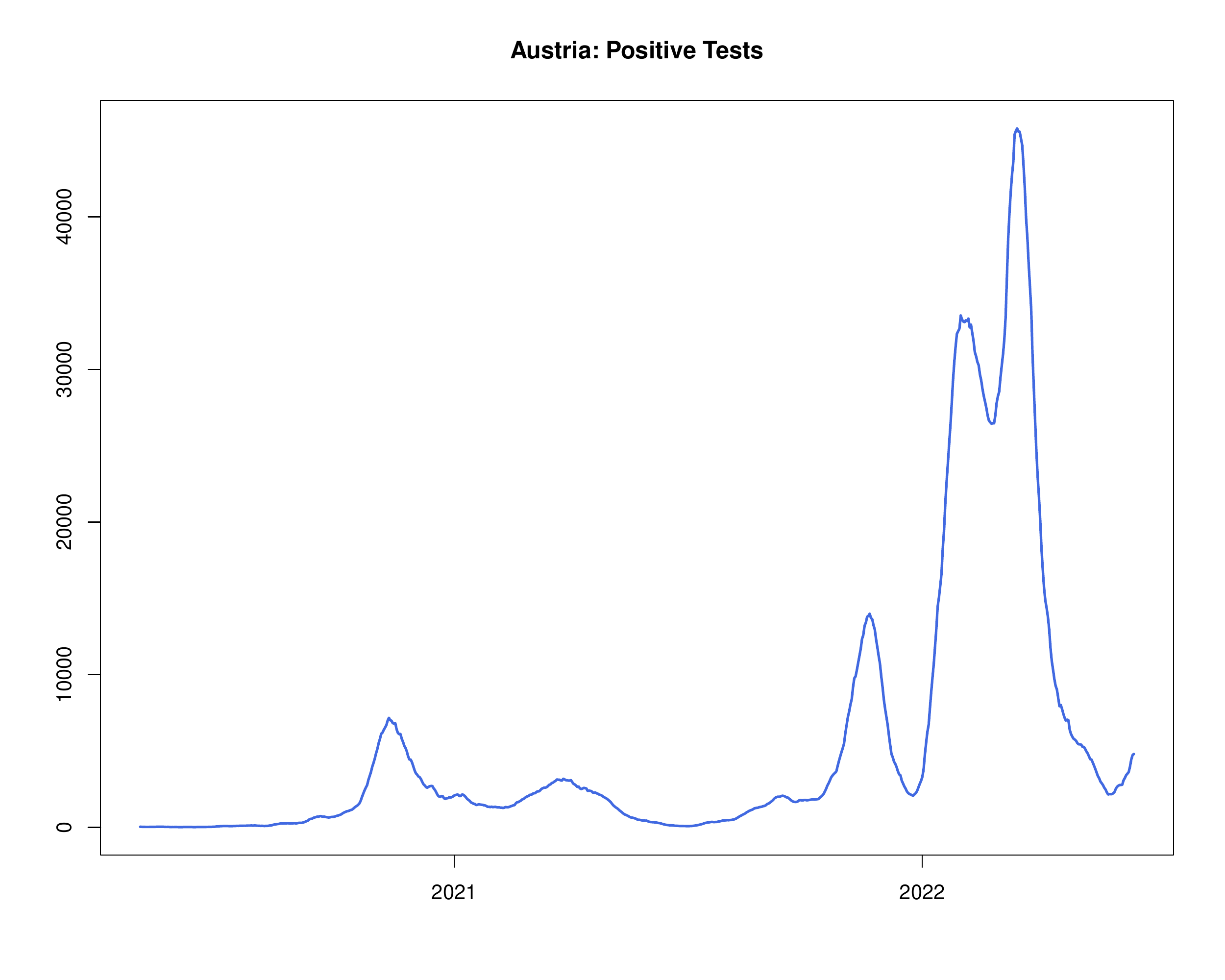}
  \caption{\footnotesize Confirmed cases of Covid-19 in Austria (from May 1, 2020 to June 15, 2022).}\label{fig:observation}
\end{figure}

In this section, we apply the nested particle filtering approach to Austrian Covid-19 data from May 1, 2020 to June 15, 2022 \footnote{The data used for this analysis are publicly available from the AGES website  https://covid19-dashboard.ages.at/}. We did not include further data, as this would have meant to consider earlier and later periods of the pandemic in which policy measures to contain the virus (such as quarantine regulations) and testing behavior of the Austrian population were substantially different.
Figure \ref{fig:observation} shows the positive tests recorded over these two years. Note that we used a seven-day rolling average of confirmed cases to avoid weekly seasonality effects, such as the fewer tests performed over the weekend. We have set the exogenous parameters  $(q,\gamma,\delta)$ and the hyperparameters of the particle filter algorithm according to the values in Appendix~\ref{app:data_parameters}.

\begin{figure}[h]
  \centering
  \includegraphics[scale = 0.42]{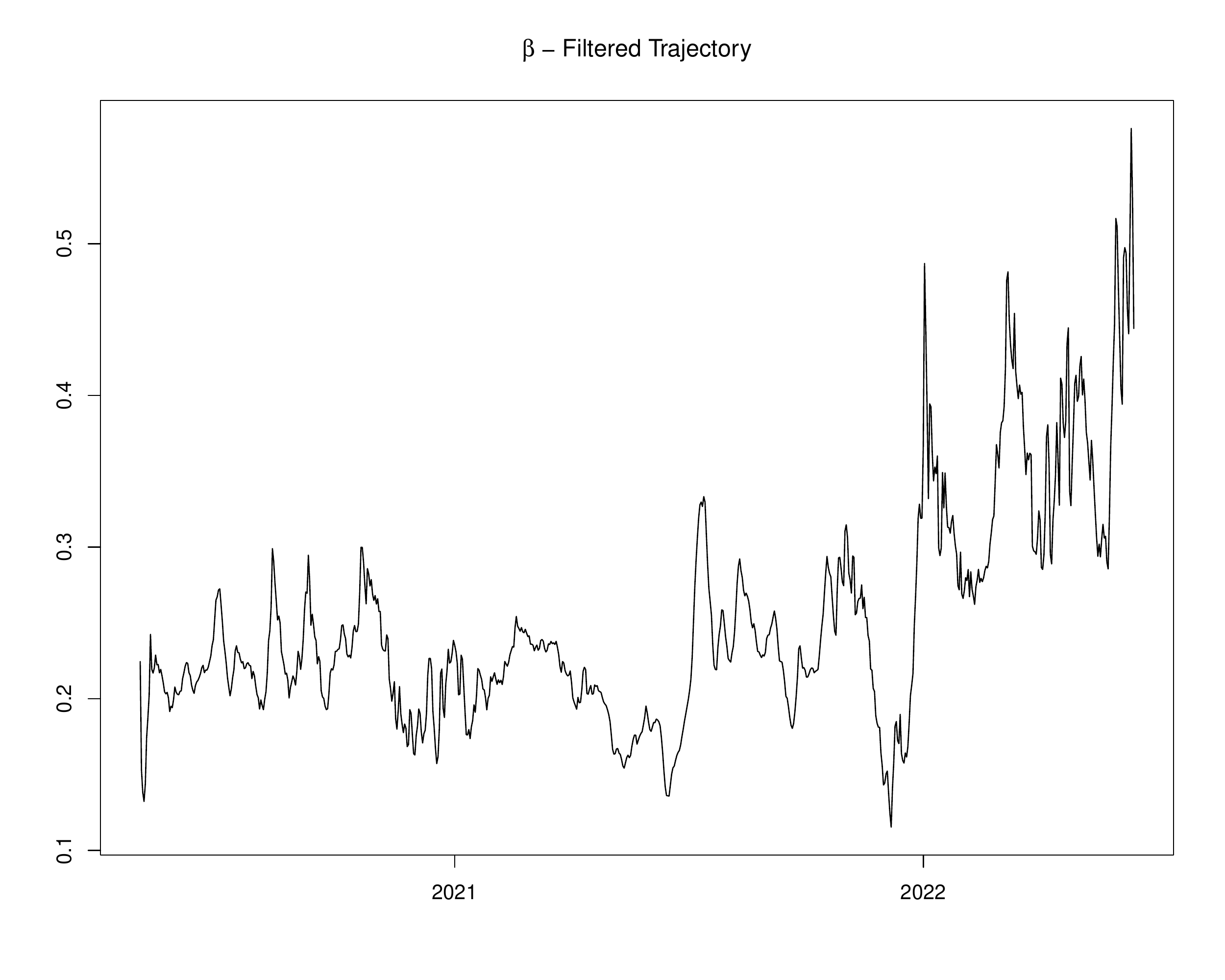} 		
  \caption{\footnotesize Estimates for the infection rate $\beta$ from Austrian Covid-19 data (from May 1, 2020 to June 15, 2022).}\label{fig:beta_data}
\end{figure}

We begin with the filtered estimate of the infection rate $(\beta_n)_{n \ge 1}$, which is provided in Figure \ref{fig:beta_data}. Here, we observe an upward trend from the beginning of 2022, which is most likely due to the arrival of an highly contagious virus variant (Omicron).

\begin{figure}[h]
\includegraphics[scale = 0.42]{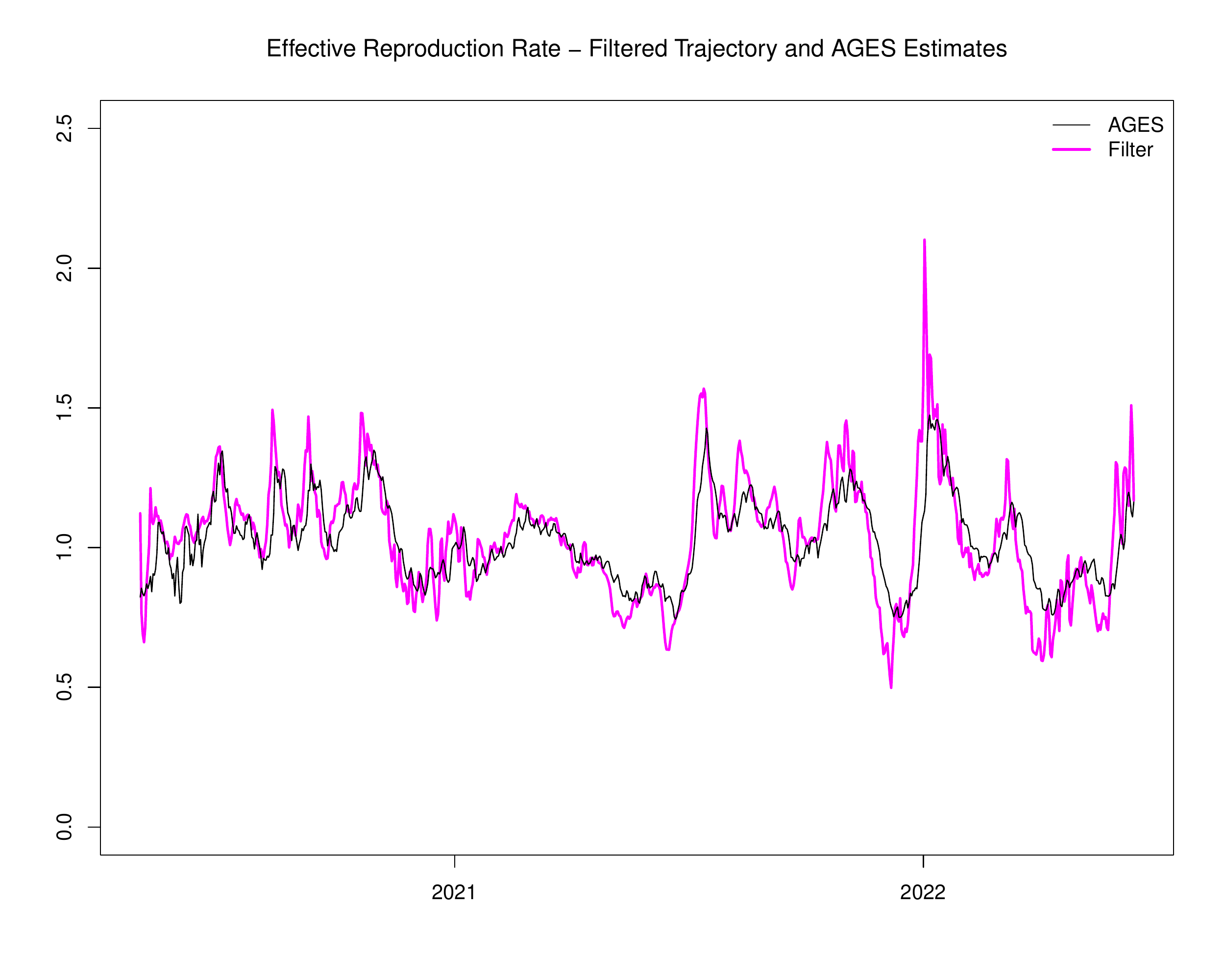}		
\caption{\footnotesize Estimates for the effective reproduction rate from Austrian Covid-19 data (from May 1, 2020 to June 15, 2022): in magenta the filtered estimate of $\Rep_n$ using our methodology; in black the estimate published by the Austrian health agency AGES.}\label{fig:Reff-data}
\end{figure}

Next, we focus on the effective reproduction rate $(\Rep_n)_{n \ge 1}$, see equation~\eqref{eq:Rep-n}. In
Figure~\ref{fig:Reff-data}, we compare our filtered estimates (in magenta) with the official estimate published by the Austrian health agency AGES (in black). The latter is computed using a simple Bayesian model with Gamma-distributed prior for $\beta$ and Poisson observations, see \cite{bib:richter-et-al-AGES} for details (see also the supplementary material of \cite{bib:cori2013new} for a description of the methodology). The plot shows that the qualitative behaviour of both estimates is very similar; however, our filtered estimate exhibits more variability and higher spikes, particularly starting from mid-2021. This seems to suggest that our filter reacts faster to changes. Note that, while the infection rate $\beta$ in the first half of 2022 is persistently higher than in 2021 (cf. Figure \ref{fig:beta_data}), the effective reproduction rate displays a spike at the beginning of 2022 and then immediately settles again around one. This is  due to the counteracting effect that a large part of the population got infected in a small time window due to higher contagiousness of the Omicron variant, which reduced substantially the number of susceptible individuals.

\begin{figure}[h]
\centering
\includegraphics[scale= 0.42]{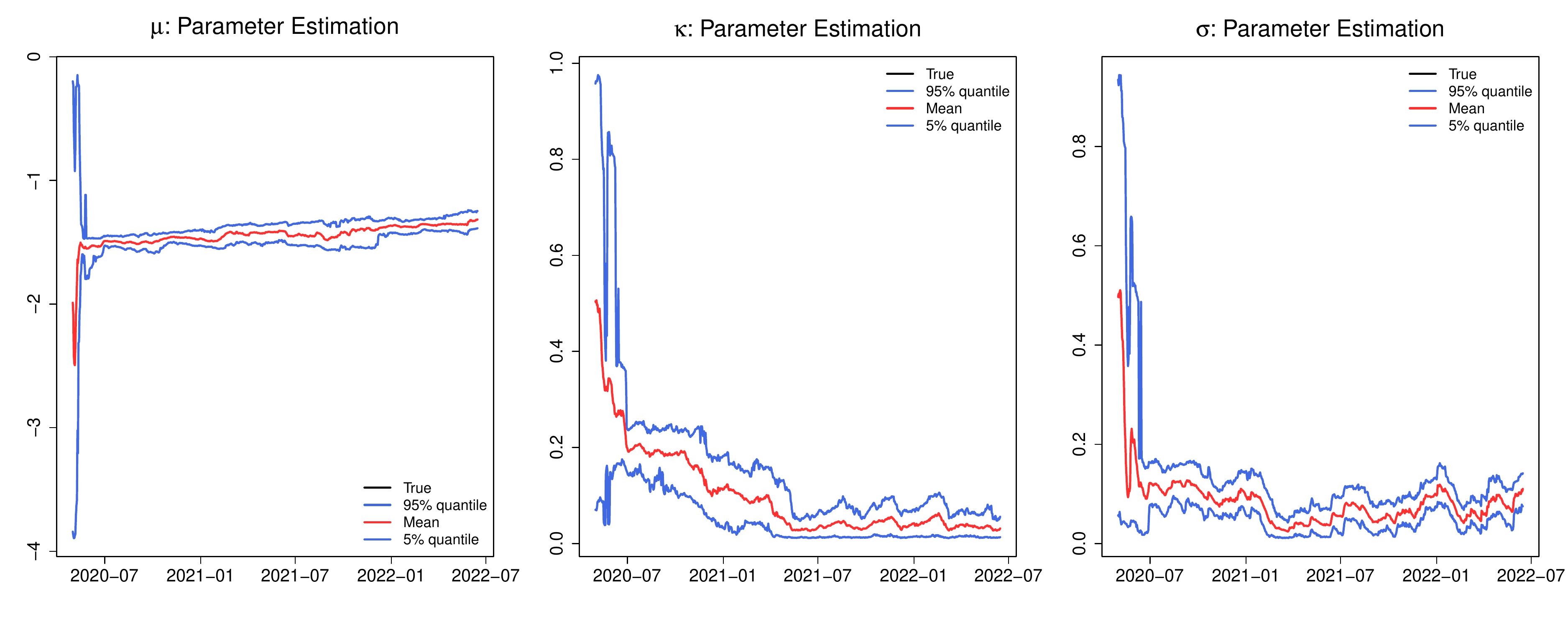}
 \caption{\footnotesize Posterior estimates for $\mu$, $\kappa$, and $\sigma$. The black line corresponds to the true value of the parameter, the two blue lines to the 5\%- and 95\%-quantile of the posterior distribution and the red line to its mean.}\label{fig:parameters-data}
\end{figure}

Figure~\ref{fig:parameters-data} shows how, over time, the posterior distribution concentrates around a specific value $\theta = (\kappa, \sigma, \mu)$. In line with the results from the simulation study, we observe that the posterior-mean estimate of the parameter $\mu$ fluctuates the least. Notice that, from the beginning of the second year of data, the posterior-mean estimate of $\mu$ seems to increase. This effect may suggest that parameters are, in reality, time-varying (in particular, the arrival of new virus variants might have started a new regime). However, we need to be careful with such conclusions, since our previously-conducted simulation study revealed the limitations in the accuracy of estimates obtained using a relatively short observation series. One possible way to investigate our conjecture  is to split the data in two periods -- each potentially corresponding to a different regime -- and perform parameter inference separately in both to compare estimates; however, this would further restrict the amount of available data, strongly worsening the accuracy and reliability of the resulting estimates and making this approach \emph{de facto} unfeasible for this case study.

\section{Forecasting and  model tests}\label{sec:predictions}

A key application of an epidemiological model is to make forecasts regarding  the development of infection numbers, which are used to gauge potential stress for the health system and which serve as a basis for decisions on containment measures. Moreover, analyzing the quality  of model-based predictions represents a natural way of testing  a given model. Therefore, in this section, we discuss  forecasts and model tests for our setup.

\subsection{Methodology} The key quantity for forecasting and testing  is the \emph{predictive distribution} of future positive tests over a time horizon $\Delta$, with  distribution function
$${F}_{n,\Delta}( x) =  \mathbb{P}(P_{n + \Delta} \le x\mid \F_{n})\,.$$
To compute an estimate $\widehat{F}_{n,\Delta} $ of ${F}_{n,\Delta}$, we rely on  simulations: we first run the particle filter over the period $[0, t_n]$, which provides an approximation of the conditional distribution of the state variables  and the  model parameters  given $\F_n$.  We then draw realisations of $I_n, R_n, \psi_n$ and of the parameters $\mu, \kappa, \sigma $  from that distribution, which we use to generate trajectories of $I,R,\psi$ and $P$ over the horizon $t_n, \dots, t_n + \Delta$ using the dynamics \eqref{eq:dynamics-1}~and~\eqref{eq:dynamics-Psi}.

There are various ways to generate \emph{point forecasts}  from the predictive distribution. It is natural to use elicitable forecasts (forecasts minimizing a suitable scoring function), such as the median, higher quantiles or the mean. Now, in our setup the predictive distribution is skewed with a very heavy upper tail  (see next paragraph) and  the mean  is  quite unstable. This suggests the use of quantile-based forecasts.\footnote{Since underestimating future infection numbers has more adverse consequences than overestimating them,  one might want to work with  higher quantiles such as the $75\%$-quantile instead of the more commonly used  median.}

We fix some horizon $\Delta$ and consider non-overlapping prediction dates $t_{n_1},  t_{n_2}, \dots, t_{n_m}$, where  $t_{n_j}= t_{n_1} + j \Delta$. Then formal \emph{statistical tests} of our methodology can be based  on the following classical result of \cite{bib:rosenblatt-52}:  if the  predictive distribution is correctly specified, that is $\widehat{F}_{n_j, \Delta} = {F}_{n_j, \Delta}$ for all $j$ (this is the null hypothesis for our model test), then,  the  random variables $\widehat{U}_j:= \widehat{F}_{n_j}(P_{n_j+\Delta} )$, $1 \le j \le m$,  are independent and identically distributed standard  uniform\footnote{%
Strictly speaking, this is true only if $\widehat{F}_{n, \Delta}$ is continuous. In our setup $P_n$ is conditionally binomial  and $\widehat{F}_{n, \Delta}$ is computed by simulation, therefore it is discrete. However, the number of simulations used is large and the conditional distribution of $P_n$ is very well approximated by a normal, so that under the null hypothesis the distribution of $\widehat{F}_{n, \Delta}(P_{n+\Delta})$ is very close to a standard uniform distribution.}.

This result is the basis for a multitude of statistical tests; for instance, see \cite{bib:gordy-mcneil-20}. Simple tests use \emph{quantile exceedances}. Fix $\alpha \in [0,1]$.  Then, the sequence of quantile exceedances
$$ I_{n_j}^\alpha = 1_{\{ P_{n_j+\Delta} > {q}_{\alpha}(\widehat{F}_{n_j,\Delta}) \}}, \quad 1 \le j \le m, $$
consists of independent and identically  Bernoulli-distributed random variables with $p = 1- \alpha$. That implies that the number of exceedances $M^\alpha = \sum_{j=1}^{m}  I_{n_j}^\alpha $  has a binomial distribution with parameters $m$ and $p=1-\alpha$, which can be tested with a simple binomial test.

More generally, one may test several quantile exceedances jointly by means of \emph{multinomial tests} as explained below (see also \cite{bib:kratz-lok-mcneil-18}). Fix quantile levels $0=\alpha_0<\alpha_1 < \dots \alpha_l < \alpha_{k+1}=1$. For $0 \le l \le k$, denote by
$$
M^{[\alpha_l,\alpha_{l+1}]} = \sum_{j=1}^{m} 1_{\{ {q}_{\alpha_{j}}(\widehat{F}_{n_j,\Delta}) \le  P_{n_j+\Delta} <  {q}_{\alpha_{j+1}}(\widehat{F}_{n_j,\Delta}) \}}
$$
the number of visits of $P_{n_j+\Delta}$ to the interval $[{q}_{\alpha_{j}}(\widehat{F}_{n_j,\Delta}), {q}_{\alpha_{j+1}}(\widehat{F}_{n_j,\Delta})]$ for all $1\le j \le m$, or, equivalently, the number of visits of  $\{\widehat{U}_{n_j}, 1\le j \le m\}$ to the cell $[\alpha_l, \alpha_{l+1})$. It follows that the $k+1$-dimensional random vector  $(M^{[\alpha_0,\alpha_{1}]}, \dots,M^{[\alpha_k,\alpha_{k+1}]})$, has a  \emph{multinomial} distribution with probabilities $p_1 = \alpha_1$, $p_2 = \alpha_2 - \alpha_1, \dots, p_{k+1} =1 -\alpha_k\,$, which can be tested with goodness of fit tests such as the exact multinomial test, see \cite{bib:EMT-package-21}.

\subsection{Empirical results}

We now apply these ideas to the Austrian Covid-19 data. We consider horizons up to 14 days, as this is a common forecasting horizon\footnote{For instance, \cite{bib:prognosis-model-austria-20} also consider prediction horizons of one and two weeks.}. The parameters of the model and the settings for the nested particle filtering algorithm are given in Appendix~\ref{app:data_parameters}.

\subsubsection*{Predictive distribution} 
\begin{figure}[h]{\centering
\includegraphics[scale = 0.45]{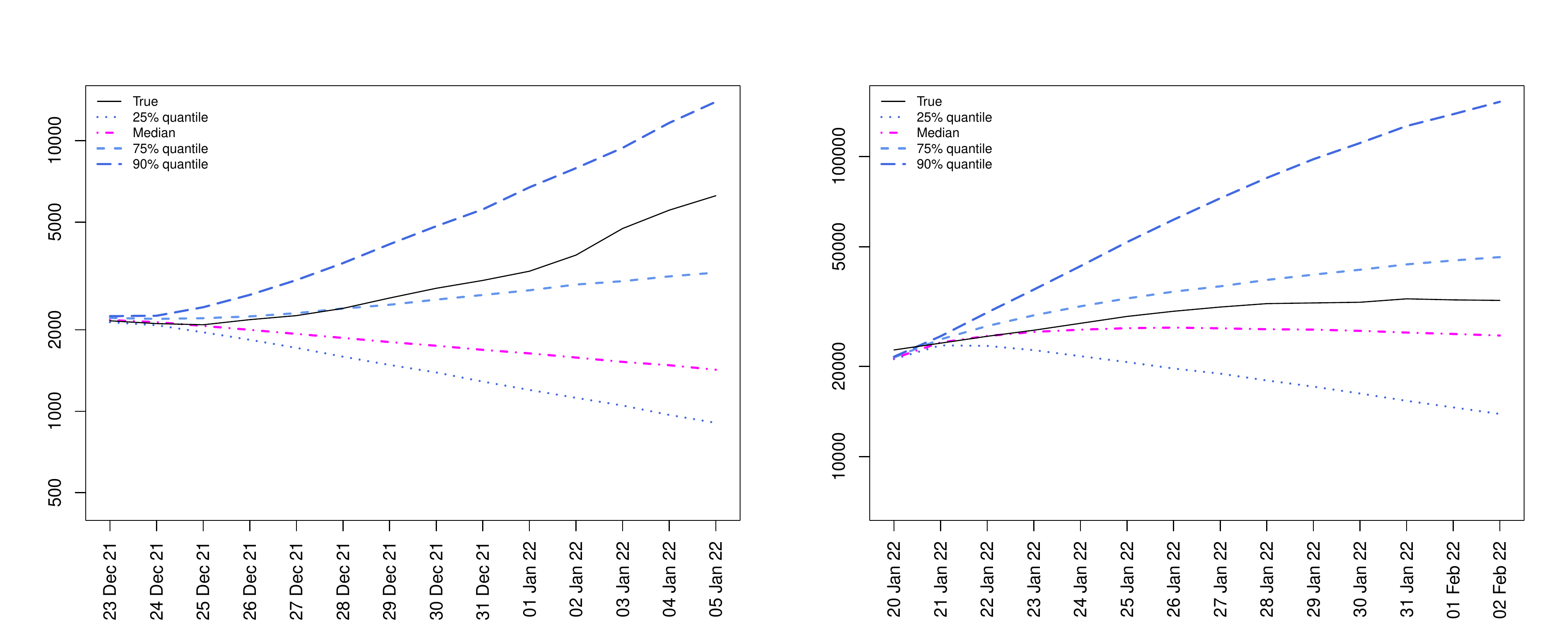}	
\caption{\footnotesize Left: quantiles $q_\alpha(\widehat{F}_{n, \Delta})$,  $\alpha =0.25,0.5, 0.75,0.9$, of $\widehat{F}_{n, \Delta}$  for the prediction date $t_n = 23/12/2021$  and prediction horizon $\Delta$ between one and 14 days together with the  actually observed  positive tests; right quantiles and mean of of $\widehat{F}_{n,\Delta}$ and realized positive tests for $t_n = 20/01/2022$.}\label{fig:predictions}
}
\end{figure}

In Figure~\ref{fig:predictions},  we plot several  quantiles of $\widehat{F}_{n,\Delta} $ for $\Delta =1,2,\dots,14$ together with the actually observed positive tests for two different prediction dates $t_n$.  The left plot shows the  forecasts made on December 23, 2021 -- that is, shortly before the start of the Omicron wave in Austria -- while the right plot shows the forecasts from January 20, 2022. Since the predictive distribution is very skewed, we use a logarithmic scale for the quantiles. The strong skewness of the predictive distribution can also be seen from Table~\ref{table:quantiles}, where we report numerical values of various  quantiles and of the mean of $\widehat{F}_{n_j, 14}$ for  these two prediction dates.

Looking at the plot on the left, we see that the model predicts the decline of the Delta wave well, but the high infection numbers caused by the onset of Omicron are close to the 90\% quantile of the predictive distribution.
This is to be expected: since our model is informed only by observations of past positive tests, it is not ``aware of'' the emergence of a new virus variant. Such information would need to be entered manually by the epidemiologist, for instance as an artificial upward shift in the distribution of the infection rate at $t_n$.
The right plot shows that by  January 20, 2022, the model has learned the different regime and actual cases are between the median and  the 75\% quantile of $\widehat{F}_{n,\Delta}$.
However, even on December 20, the infection numbers of the  Omicron wave are below the 90\% quantile of the predictive distribution, and thus  well  within the range of possible future scenarios generated by our model.

\begin{table}
\begin{tabular}{ l cccccccc}
\hline
Prediction Date    && 10\%  &  25\%    &  50\%    &  75\%    &  90\%    &   95\%     & Mean  \\ \hline
Dec 23, 2021  && 494   &  851     & 1411     &  3509    &  16338   &   79142    & 20194 \\
Jan 20, 2022  && 8388  &  13372   &  24845   &  46603   &  155557  &  248978    & 56468 \\ \hline
\end{tabular}

\vspace{0.1cm}
\caption{\footnotesize Quantiles and mean of the predictive distribution  $\widehat{F}_{n, 14}$ for two different prediction  dates and a horizon of 14 days. Note that the distribution is very skewed with high upper quantiles and a mean exceeding  the the 90\%  quantile (for December 23, 2022) respectively  the 75\% quantile (for January 20, 2022).}\label{table:quantiles}
\end{table}

\subsubsection*{Formal tests.}  For our tests, we use a horizon of $\Delta = 14$. Since posterior-mean estimates resulting from the nested particle filtering algorithm start stabilising after a little more than a year, we  chose  September 1, 2021, as the date of the first test, which yields  $m = 20$ non-overlapping testing dates. We consider the quantile levels $\alpha_1=0.25$, $\alpha_2 =0.5$, $\alpha_3= 0.75$, $\alpha_4 = 0.9$ in our tests. In Table~\ref{table:exceedances}, we report the expected  and the observed  number of quantile exceedances and cell visits. We see that the quantile exceedances are quite close to their expected value, and a binomial test applied to the observed quantile exceedances yields high $p$-values (details are omitted). Finally, we ran the exact multinomial test for cell visits  and obtained a $p$-value of 0.521. These test results provide support for the methodology proposed in this paper. Note, however, that
the precise outcome of numerical tests varies somewhat with the chosen quantile levels, testing horizon and testing dates.  Moreover, the estimated predictive distribution $\widehat{F}_{n,\Delta}$ is sensitive  to the choice of settings in the nested particle filter, and that some experimentation is therefore necessary (especially given the relatively short series of observations).

\begin{table}\label{table:exceedances}
\begin{tabular}{ l cccc}
\hline
$\alpha$ & 0.25 &  0.5  &  0.75  &  0.9  \\ \hline
exp. & 15   &  10   &   5    &   2   \\
obs. & 15    & 11   &   4    &   0   \\
\hline
\end{tabular}\hspace{1cm}
\begin{tabular}{ l ccccc}
\hline
cell & < 0.25    & 0.25-- 0.5  & 0.5--0.75   & 0.75--0.9  & >0.9  \\ \hline
exp. &  5        &  5          & 5           &      3     &    2  \\
obs. &  5        &  4          & 7           &      4     &    0  \\
\hline
\end{tabular}
\vspace{0.1cm}
\caption{\footnotesize Left: expected and observed quantile exceedances; right: expected and observed cell visits.}
\end{table}

\subsection*{Acknowledgements}
{We are grateful to Sylvia Fr\"{u}hwirth-Schnatter, Luca Gonzato and Giorgia Callegaro for useful comments and suggestions. The work of Katia Colaneri was partially supported by project PRIN 2022 ``Stochastic control and games and the role of information'' (2022BEMMLZ, National PI T. De Angelis, Local Investigator K. Colaneri) funded by the Italian Ministry of University and Research, MUR. Katia Colaneri is also member of Indam-Gnampa. The work of Camilla Damian was partially supported by the Austrian Science Fund project ``Detecting gender bias in children's books''(FWF 1000 Ideas Programme TAI 517-G, PI L. Vana G\"ur).} 

\subsection*{Conflict of Interest}
The authors have no conflicts of interest to disclose.

\appendix

\section{Details on the nested particle filtering}\label{app:particle-filter}


\subsubsection*{Non-Integer Quantities.} Note that some quantities in the model described by \eqref{eq:dynamics-1} are not integers due to the presence of rates $\gamma > 0$ and $\delta > 0$, while others (e.g. $P$ and $I^+$) are nonnegative integers by construction. In particular, it might be that the process $I$, i.e. the number of infectious people, gets below one (or even becomes negative) for some particles at some point in time. To avoid this issue we artificially set $I_n=0$, whenever a particle is propagated to a negative value.
Notice that $P_n$ could be a positive integer even if for some particles  $\lfloor I_n\rfloor = 0$. In this case the likelihood of the observation given the particle is zero and hence the particle is eliminated in the resampling.

\subsubsection*{Computational Details.} All computations for this paper are done using \textsf{R} (see \cite{R2022}). The triple of latent state variables $(I, R, \Psi)$ is stored as a $K \times M \times 3$ array object, where rows correspond to parameter particles, columns to state particles and each $K \times M$ matrix slice along the third dimension corresponds to one of the three state variables. This structure makes the code shorter, readable and relatively efficient, as vectorized operations in \textsf{R} could be used for instance for the state particle evolution. Most importantly, resampling steps can be performed simultaneously along given dimensions of the array.

Although the code is not optimized for speed, and some filtering-specific choices as resampling at each iteration, or jittering every parameter particle, are time consuming, it is relatively fast. One round of the nested particle filter for our application, considering 365 time steps (i.e., one year of data), takes around 1.4 minutes (on the processor AMD Ryzen 7 PRO 4750U).

As computing an estimate for mean relative errors requires averaging over independent runs of the algorithm, we use the package \textsf{parallel}; in particular, we use the parallelized analogous of \textsf{lapply}, \textsf{mclapply}, on 5 cores. \footnote{Note that \textsf{mclapply} is not available for Windows, except in the serial sense -- that is, resulting in a call to \textsf{lapply}, see for instance https://stat.ethz.ch/R-manual/R-devel/library/parallel/doc/parallel.pdf.}
Overall, it takes about 26 minutes for 50 independent runs of the algorithm for one year of data.

\section{Inputs for the simulation study}\label{app:simulation}

\paragraph{Simulation parameters (true values) and other inputs}
\begin{itemize}
 \item $q = 10\%$.
 \item $\gamma= 1/10$.
 \item $\delta=1/200=0.05$.
 \item $\kappa=0.2$.
 \item $\sigma=0.1$.
 \item $\mu=\log(\gamma+q)-\frac{\sigma^2}{2\kappa}\approx-1.63$.
 \item Number of individuals: $N=8.917 \cdot 10^6$. 
\item Number of days:  $N^{days} = 731$ (including time 0).
\item Time step: $\Delta=t_n-t_{n-1}= 1$ day.
\item $R_0=0$.
\end{itemize}
\paragraph{Settings for the nested particle filtering}
  \begin{itemize}
  \item Prior for $\Psi_0$: normal with mean $\mu$ and standard deviation $0.175$. 
  \item Prior for $I_0$: gamma (to ensure non-negativity) with mean $3000$ and variance $15000$. 
  \item Prior for $\kappa$, $\sigma$ and $\mu$: uniform over $1\%$ and $500\%$ of true values.
  \item Number of particles in the state space $M=500$ and in the parameter space $K=500$.
  \item Variance of the jittering kernels $\epsilon_\kappa=\epsilon_\mu=\epsilon_\mu= 5K^{-2}$.
\end{itemize}

\section{Inputs for the real data analysis}\label{app:data_parameters}
\paragraph{Data characteristics}
  \begin{itemize}
  \item Time period: from May 1, 2020, to June 15, 2022.
  \item Total number of individuals: $N = 9.028 \cdot 10^6$ (i.e. population of Austria).
  \item Observations $P$: 7-day rolling average of confirmed cases.
\end{itemize}
\paragraph{Fixed parameters and other inputs}
  \begin{itemize}
  \item $q = 10\%$.
 \item $\gamma= 1/10$.
 \item $\delta=1/200=0.05$.
\item Number of days:  $N^{days} = 776$ (including time 0).
\item Time step: $\Delta=t_n-t_{n-1}= 1$ day.
\item $R_0=0$.
\end{itemize}

\paragraph{Settings for the Nested Particle Filtering}
  \begin{itemize}
  \item Prior for $\Psi$: normal with mean $\log(\gamma+q) \approx -1.61$ and standard deviation $0.1$.
  \item Prior for $I_0$: gamma (to ensure non-negativity) with mean $470$ and variance $2350$, where the mean corresponds to $\bar{p}/q$ and $\bar{p} = 47$ to the 7-day rolling average of confirmed cases on the day preceding the start of our data analysis (i.e. on April 30, 2022).
  \item Prior for $\kappa$: uniform over $[0.01, 1]$. This prior is chosen to be uninformative.
  \item Prior for $\sigma$: uniform over $[0.01, 1]$ (uninformative).
  \item Prior for $\mu$: uniform over $[-4,-0.01]$ (uninformative).
  \item Number of particles in the state space $M = 600$ and in the parameter space $K = 600$.
  \item Variance of the jittering kernels $\epsilon_\kappa=\epsilon_\mu=\epsilon_\mu= 5K^{-2}$.
\end{itemize}

\end{document}